\documentclass[11pt,letter]{emulateapj}
\usepackage{color}
\usepackage{amsmath,amssymb}
\usepackage{natbib,longtable}

\newcommand{\si}{Si{\sc{ii}}~$\lambda$4000}
\newcommand{\sifive}{Si{\sc{ii}}~{$\lambda$}5800}
\newcommand{\sisix}{Si{\sc{ii}}~{$\lambda$}6150}
\newcommand{\mg}{Mg{\sc{ii}}~$\lambda$4300}
\newcommand{\otio}{{\"O}10}
\newcommand{\ntio}{N10}
\newcommand{\dpew}{$\Delta$pEW}
\newcommand{\salttwo}{{\sc{SALT2}}}
\def\lsim{\raise0.3ex\hbox{$<$}\kern-0.75em{\lower0.65ex\hbox{$\sim$}}}
\def\gsim{\raise0.3ex\hbox{$>$}\kern-0.75em{\lower0.65ex\hbox{$\sim$}}}

\newcommand{\samplenbr}{55}
\newcommand{\sampleref}{26}
\newcommand{\samplesdss}{16}
\newcommand{\samplesnls}{13}
\newcommand{\samplegrad}{22}

\newcommand{\sampledivide}{3}
\newcommand{\noredccorr}{5.2}
\newcommand{\ccorr}{5.8}
\newcommand{\gradcorr}{3.8}
\newcommand{\noredgradcorr}{4.3}
\newcommand{\pewepochk}{0.3}
\newcommand{\scaleoverlap}{64}
\newcommand{\sicorrall}{0.67}

\newcommand{\comment}[1]{}

\newcommand{\xclude}[1]{} % exclude for now
\newcommand{\fixme}[1]{}
\newcommand{\changed}[1]{#1}
\newcommand{\new}[1]{#1}

\newcommand{\newnew}[1]{#1}

\newcommand{\linda}[1]{#1}

\newcommand{\refcom}[1]{#1}

\newcommand{\comreply}[1]{#1}

\begin{document}

%\title{The {\si} feature of Type Ia supernovae and correlations with lightcurve color, subclasses and expansion velocity}

\title{Evidence for a correlation between the {\si} width and Type Ia supernova color}

\author{J. Nordin\altaffilmark{1,2}, L. \"{O}stman\altaffilmark{1,3}, A. Goobar\altaffilmark{1,2},
C.~Balland\altaffilmark{4,5}, H. Lampeitl\altaffilmark{6}, R.~C. Nichol\altaffilmark{6}, M. Sako\altaffilmark{7}, D.~P. Schneider\altaffilmark{8}, M. Smith\altaffilmark{9}, J. Sollerman\altaffilmark{2,10} and J.~C. Wheeler\altaffilmark{11}
}

\altaffiltext{1}{Department of Physics, Stockholm University, 106 91 Stockholm, Sweden; nordin@physto.se}
\altaffiltext{2}{Oskar Klein Centre for Cosmo Particle Physics, AlbaNova, 106 91 Stockholm, Sweden}
\altaffiltext{3}{Institut de F\'{i}sica d'Altes Energies, 08193 Bellaterra, Barcelona, Spain}
\altaffiltext{4}{LPNHE, CNRS/IN2P3, Universit\'e Pierre et Marie Curie Paris 6, Universit\'e Denis Diderot Paris 7, 4 place Jussieu, 75252 Paris Cedex 05, France}
\altaffiltext{5}{Universit\'e Paris 11, Orsay, F-91405, France}
\altaffiltext{6}{Institute of Cosmology and Gravitation, Portsmouth PO13FX, UK}
\altaffiltext{7}{Department of Physics and Astronomy, University of Pennsylvania, Philadelphia, PA 19104, USA}
\altaffiltext{8}{Department of Astronomy and Astrophysics, Pennsylvania State University, University Park, PA 16802, USA}
\altaffiltext{9}{Astrophysics, Cosmology and Gravity Centre (ACGC), Department of Mathematics and Applied Mathematics, University of Cape Town, Rondebosch 7700, South Africa}
%\altaffiltext{10}{Dark Cosmology Centre, Niels Bohr Institute, University of Copenhagen, DK-2100, Denmark.}
\altaffiltext{10}{Astronomy Department, Stockholm University, AlbaNova University Center, 106 91 Stockholm, Sweden}
\altaffiltext{11}{Department of Astronomy and McDonald Observatory, The University of Texas, 1 University Station C1402, Austin, TX 78712-0259, USA}

%\maketitle

% Abstract
\begin{abstract}
We study the pseudo equivalent width of the {\si} feature of Type Ia
supernovae (SNe Ia) \changed{in the redshift range $0.0024 \le z \le
0.634$}. We find that \changed{this spectral indicator} correlates
with the light curve color excess (\salttwo~c) as well as previously
defined spectroscopic subclasses (Branch types) and the
evolution of the {\sisix} velocity, i.e., the so called velocity
gradient.  \changed{Based on our study of \samplenbr \ objects from
different surveys, we find indications that the {\si} spectral indicator
could provide important information to improve cosmological distance
measurements with SNe Ia.}
\end{abstract}

\keywords{cosmology: observations -- distance scale -- dust,
extinction -- supernovae: general}

\section{Introduction}

Type Ia supernovae (SNe Ia)
%, as standard candles, 
provided the first evidence for the accelerating expansion rate of the
universe \citep{1998AJ....116.1009R,1999ApJ...517..565P}. This \newnew{was} made possible by the homogeneous luminosities of this class of explosions, which \newnew{were} further ``standardized'' using two empirically
derived relations: SNe with wider
light curves are brighter \citep{1993ApJ...413L.105P} and redder SNe
are fainter \citep{tripp98}. The latter effect is often attributed
as mainly due to extinction by interstellar dust in the host
galaxy.  In recent years, however, several new results have
complicated the ``simple'' view that the peak luminosities of
normal SNe Ia can accurately be described using only two empirical
parameters, light curve width and color.

The measured Doppler shift of the {\sisix} feature, the main
defining characteristics of SNe Ia, is the most straightforward
estimate of the expansion velocity \citep[\linda{but possibly not the
most accurate, see}][]{1996MNRAS.278..111P}.
\citet{2005ApJ...623.1011B} studied \new{the change of {\sisix}
velocity} with time, the \emph{velocity gradient}, and used this
quantity to classify SNe as high velocity gradient (HVG), low velocity
gradient (LVG) or faint. HVG and LVG SNe have similar light curve
widths but different expansion evolution, thus showing that SNe Ia
cannot be completely described by the light curve width.

\citet{2010Natur.466...82M} argued that the velocity gradient is
related to shifts in nebular line velocities and that this can be
explained through viewing angle effects from a non-symmetric
explosion. \new{Signs of asymmetries can also be found through
spectropolarimetry \citep{2010arXiv1008.0651M}}.

As a further example of the diversity of SNe Ia,
\citet{2006PASP..118..560B} introduced a classification scheme based
on the shape of the {\sifive} and {\sisix} features: normal ``Core
Normal'' (CN), broader ``Broad Line'' (BL), weaker ``Shallow Silicon''
(SS) or with the deep absorptions of fainter SNe Ia, ``Cool''.
%normal ``Core Normal'' (CN), broader ``Broad Line'' (BL), weaker ``Shallow Silicon'' (SS) or with the deep absorptions of fainter SNe Ia (Cool).
%

\newnew{The photometric lightcurves of SNe Ia from different environments also exhibit systematic diversity.}
The distribution of SN Ia decline rates (light curve width) varies with host galaxy type \citep{1996AJ....112.2391H} and redshift \citep{2007ApJ...667L..37H}.
These results can be compared with the rates of SNe Ia that seem to be best explained through two populations, one ``prompt'' and one ``delayed''\citep{2005A&A...433..807M,2005ApJ...629L..85S}.
SNe in different host galaxy types also appear to have different intrinsic \newnew{light curve} properties, beyond what can be corrected using light curve width and color \citep{2010ApJ...715..743K,2010arXiv1003.5119S,2010arXiv1005.4687L}.
\citet{2009ApJ...699L.139W} compared extinction laws (fits of
total-to-selective \linda{extinction ratio}, $R_V$), finding different relations for
high and low {\sisix} velocity objects. \citet{2010arXiv1006.4612L} reported
that high velocity (often also HVG) SNe predominantly occur in Sb--Sc galaxies.
\refcom{Signs of spectroscopic evolution with redshift were claimed by \citet{2008ApJ...684...68F} and \citet{2009ApJ...693L..76S}, both reporting shallower absorption features at higher redshifts based on studies of composite spectra (where host contamination removal was attempted after and before combination, respectively).}

The extinction of SNe Ia, inferred from observations of their colors, displays a steeper law (lower $R_V$) than the typical Milky Way extinction law \citep{2006AJ....131.1639K,2006MNRAS.369.1880E,guy07,2007AJ....133...58K, nobili07,
2008MNRAS.384..107E,2008ApJ...675..626W}, and is at least for some objects well
fitted with simple models assuming circumstellar dust \citep{2008ApJ...686L.103G, 2010AJ....139..120F}.

Precision cosmology with SNe Ia depends critically on the
homogeneity of the SN brightness.  Potentially, further
sharpening of the ``standard candle'' could be accomplished using
spectroscopic properties or near-IR SNe Ia data
\citep{2009arXiv0905.0340B,2010AJ....139..120F}. At the same time,
systematic effects caused by poorly corrected brightness evolution and
reddening limit current and future SN Ia dark energy studies
\citep{2008JCAP...02..008N}. 
Thus, characterizing subclasses and empirical secondary relations is essential for the use of SNe Ia to obtain precise constraints on e.g., the nature of dark energy.

In \citet[][, hereafter N10]{2011A&A...526A.119N}, we performed
an analysis of individual spectra observed at the \linda{New Technology Telescope (NTT) and the Nordic Optical Telescope} \citep[NOT;][hereafter \otio]{2011AA...526A..28O} in conjunction with the SDSS-II Supernova Survey
\citep{1998AJ....116.3040G,2000AJ....120.1579Y,2008AJ....135..338F,2008AJ....135..348S}. 
The NTT/NOT spectroscopic sample is publicly available, see {\otio}~for a full description.
These \newnew{spectra}
were compared with a large set of nearby SN Ia spectra. Pseudo~equivalent widths (pEWs) and line velocities were measured for \newnew{selected}
optical features of normal SN Ia spectra. We examined
possible correlations between SN Ia spectral indicators (over a wide
range of epochs) and redshift, SALT \citep{guy05} light curve
parameters and host galaxy properties. 
The {\si} feature around light curve peak was found to correlate strongly
with the light curve width (stretch): SNe with wider light curves have
narrow/weaker {\si} absorption features. 
This relationship was introduced indirectly as the {\mg} ``breaking point'' in \citet{2007AA...470..411G} and explicitly as a function of {\si} width in \citet{2008A&A...477..717B}. Further discussions can be found in \citet{2008AA...492..535A} and \citet{2011MNRAS.410.1262W}.

Interestingly, \newnew{in \ntio} we also identified a tentative correlation between the
{\si} pEW and the fitted SALT color. \newnew{This correlation was particularly} strong in the epoch range $0$--$8$ days after $B$-band maximum.
The width of the small {\si} feature is only marginally affected by
interstellar dust or flux normalization (see N10). Instead, \newnew{a change} in
the pEW is likely related to photospheric temperature or elemental
abundance differences.
A correlation between feature width and light curve color would thus
\changed{indicate} that at least part of the reddening, as described by
light curve color, has an origin in or close to the explosion.
Given the potential importance of these findings, in this paper we perform further tests on the {\si} pEW using an updated light curve fitter \citep[\salttwo;][]{guy07} and an increased sample of SN Ia spectra from the SuperNova Legacy Survey (Very Large Telescope) \citep[SNLS VLT;][]{2009AA...507...85B}. In addition to light curve parameters, we also
include comparisons with velocity gradient and Branch type.

This paper is organized as follows. In Section~\ref{sec:sample} we
present the data and measurements used and in Section~\ref{sec:si} we
show our main results concerning the {\si} feature together with tests
for systematic effects. In Section~\ref{sec:dis} we discuss these
results in a wider context, and we conclude in
Section~\ref{sec:conc}. The data used are presented in Table
2. \changed{All quoted spectral epochs refer to the rest frame of the
SN and are expressed as days relative to time of peak
$B$-band luminosity.}

\section{The spectral sample and SN properties}
\label{sec:sample}

The sample used in \ntio~was a combination of SDSS-NTT/NOT spectra (\otio) and local spectra.\footnote{Many assembled using the SUSPECT data base, \texttt{http://bruford.nhn.ou.edu/suspect/}.} That sample is here enlarged in size and redshift
range by including the public SNLS VLT spectra \citep{2009AA...507...85B}. NTT/NOT SNe with host galaxy contamination above \linda{$10\%$ in the $r$ band have been} host subtracted using a Principal
Component Analysis technique (presented in detail in {\otio} and with
tests on effects on spectral indicators presented in \ntio). For the SNLS
spectra, the host galaxy light have been subtracted using the PHASE 2d-reduction pipeline \citep{2008A&A...491..567B}.\footnote{\new{In some cases the SN location was too close} to the center of the host galaxy \new{for an efficient PHASE reduction. For these cases a one-dimensional reduction was performed, see \citet{2008A&A...491..567B} for details.}} 
In this analysis, \linda{we study confirmed normal SNe Ia, removing} all SNe that are clearly peculiar (of SN~1991bg, SN~1991T, or SN~2002cx type). Furthermore,  we restrict the sample to SNe with at least one spectrum between 0 and 8 rest frame days from light curve peak. 

The spectral properties are compared with light curve parameters estimated using
the {\salttwo} light curve fitter \citep{guy07}. \new{SALT2 characterizes SN lightcurves through the \emph{peak magnitude}, \emph{x1} (roughly corresponding to light curve width) and \emph{c}. The \emph{c} parameter describes an empirically fit wavelength-\newnew{dependent} flux attenuation. SNe with high $c$-values are ``redder'' in the sense of having less flux at short wavelengths, while SNe with small $c$-values can be called ``bluer''.}
The SALT2 parameters are obtained from either the SNLS 3 yr compilation \citep{2010arXiv1010.4743G}, the Union2 sample \citep{2010ApJ...716..712A}, or for NTT/NOT SNe from our own \salttwo~fits to SDSS-II photometry \citep[derived using the method presented in ][]{2008AJ....136.2306H}. For a few local SNe, we use {\salttwo} values from \citet{2008AA...492..535A}. SNLS 3 yr light curve parameters rely on an updated version of \salttwo, we therefore rescale SNLS {\salttwo} parameters to match the {\scaleoverlap} SNLS SNe that overlap with Union2. The uncertainty of this scaling procedure is propagated into the light curve parameter uncertainties.\footnote{\linda{The added uncertainties for the rescaled light curve parameters are $\sigma_m^{\rm{rescale}} = 0.04$, $\sigma_{x_1}^{\rm{rescale}} = 0.42$ and $\sigma_{c}^{\rm{rescale}} = 0.05$.}}
SNe without good \salttwo~fits are removed from the analysis. We remove any SNe flagged as uncertain and demand a clearly defined light curve peak.

\newnew{We here discuss the {\si} feature of SNe Ia. This region is \emph{partly} created by absorption by the Si~{\sc{ii}}~$\lambda$4130~line, but with the term {\si} we also include absorption by other elements at these wavelengths.}
The {\si} pEW is estimated using the
measurement pipeline introduced in \ntio. We use the standard
formula:
\begin{equation}
  pEW = \sum_{i=1}^{N} \left( 1-\frac{f_{\lambda}(\lambda_i)}{f_{\lambda}^{c}(\lambda_i)} \right) \Delta\lambda_i,
  \label{eq:pew}
\end{equation}
where $f_{\lambda}(\lambda_i)$ is the observed flux and $f^c_{\lambda}(\lambda_i)$ is the pseudo-continuum. For
{\si}, the pseudo-continuum is defined as the line between the peak at $3800$--$3950$ {\AA} and
the peak at $4000$--$4200$ {\AA} and the sum is made over all
wavelength bins between the peaks. \linda{If both peaks are not found}, the measurement is discarded. 
\new{The minimum signal-to-noise ratio (S/N, in $25$ {\AA} bins) for accurate {\si} measurements is $\sim$8. This noise cut removes slightly more than half of both the NTT/NOT and SNLS} spectra. Statistical and systematic pEW uncertainties are estimated as in \ntio.

To remove the (weak) epoch dependence of the {\si} pEW within the narrow 
8 day window, we fit a linear function to the
{\si} pEW values as a function of spectral epoch and subtract this from
all measurements.\footnote{The line fit is done using the complete local sample, see N10. The {\si} pEW epoch dependence is \pewepochk~{\AA}~day$^{-1}$.}  
%

%\begin{equation}
%\Delta pEW = pEW - m - k \cdot  epoch
%\end{equation}
%\refcom{where a fit to local SNe yield $m = 16.66$ and $k = 0.3236$ for {\si} and $epoch$ is the rest frame epoch of the spectrum (see N10).}
%
%
After subtraction we thus have a pEW difference, $\Delta$pEW. SNe with a negative $\Delta$pEW have smaller than average {\si} and SNe with positive values have a wider absorption feature. 
\refcom{In Figure~\ref{fig:pewsample} we show examples of  {\si} features for SNe with narrow/wide {\si} and in Figure~\ref{fig:deltasample} we show pEW measurements for the sample and how these are converted to $\Delta$pEW.}
%. Figure 8 in {\ntio} further shows directly how pEW values change with epoch.

%\xclude{
{
\begin{figure}[htbp]
  \centering
  \includegraphics[angle=-90,width=0.99\columnwidth]{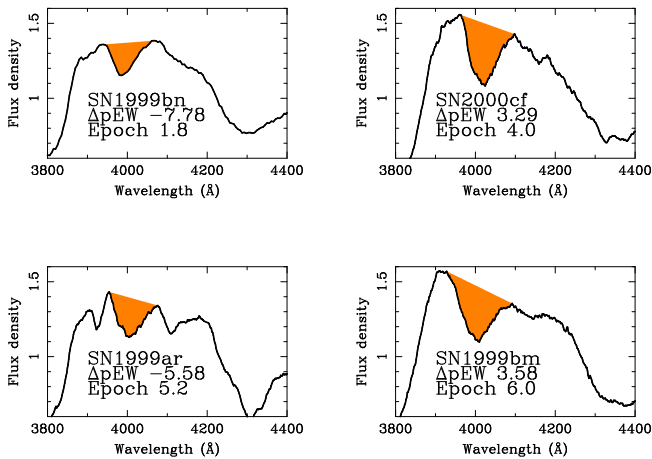}
  \caption{ {\it The {\si} feature of SNe Ia}. Top left: SN1999bn at
  day +2 ({\si} $\Delta$pEW$=-7.78$~{\AA}). Top right: SN2000cf at day +4 ($\Delta$pEW$=3.29$~{\AA}). Bottom left: SN1999ar at day +5 ($\Delta$pEW$=-5.58$~{\AA}). Bottom right: SN1999bm at day +6 ($\Delta$pEW$=3.58$~{\AA}). Left panels show SNe with narrow {\si} feature also after
  removing epoch dependence, while right panel SNe have wider Si~{\sc{ii}}~$\lambda$~4000. \newnew{Stated} \linda{epochs and} \newnew{$\Delta$pEW measurements are for these \emph{individual} spectra, and do not correspond to values for SNe with multiple spectra, which is given in Table 2.} }
  \label{fig:pewsample}
\end{figure}
}

\xclude{
\begin{figure}[htbp]
  \centering
  \includegraphics[angle=-90,width=0.99\columnwidth]{si_random_iv.ps}
  \caption{ {\it The {\si} feature of SNe Ia}. The left panel shows a selection of SN spectra of SNe with narrow {\si} regions, the right panel a (random) selection of SN spectra with wider {\si}. The spectra are ordered in increasing restframe epoch from bottom. For each spectrum the {\si} region has been shaded grey and ID, restframe epoch and $\Delta pEW$ have been stated.}
  \label{fig:pewsample}
\end{figure}
}

\xclude{
\begin{figure}[htbp]
  \centering
  \includegraphics[angle=0,width=0.99\columnwidth]{origin_pew2.eps}
  \includegraphics[angle=0,width=0.99\columnwidth]{pew2dev.eps.eps}
  \caption{ \refcom{{\it pEW and $\Delta pEW$ measurements of {\si}}. The top panel shows the pEW measurements as a function of restframe spectral epoch. The dashed line is the best fit pEW evolution derived using the full sample of N10. The lower panel shows $\Delta pEW$ measurements after subtraction of the mean epoch evolution.}
    %\refcom{SDSS NTT/NOT measurements are shown as (black) crosses, CfA as (magenta) stars, SUSPECT as (cyan) boxes and ...} 
}
  \label{fig:deltasample}
\end{figure}
}

\begin{figure}[htbp]
  \centering
  \includegraphics[angle=0,width=0.99\columnwidth]{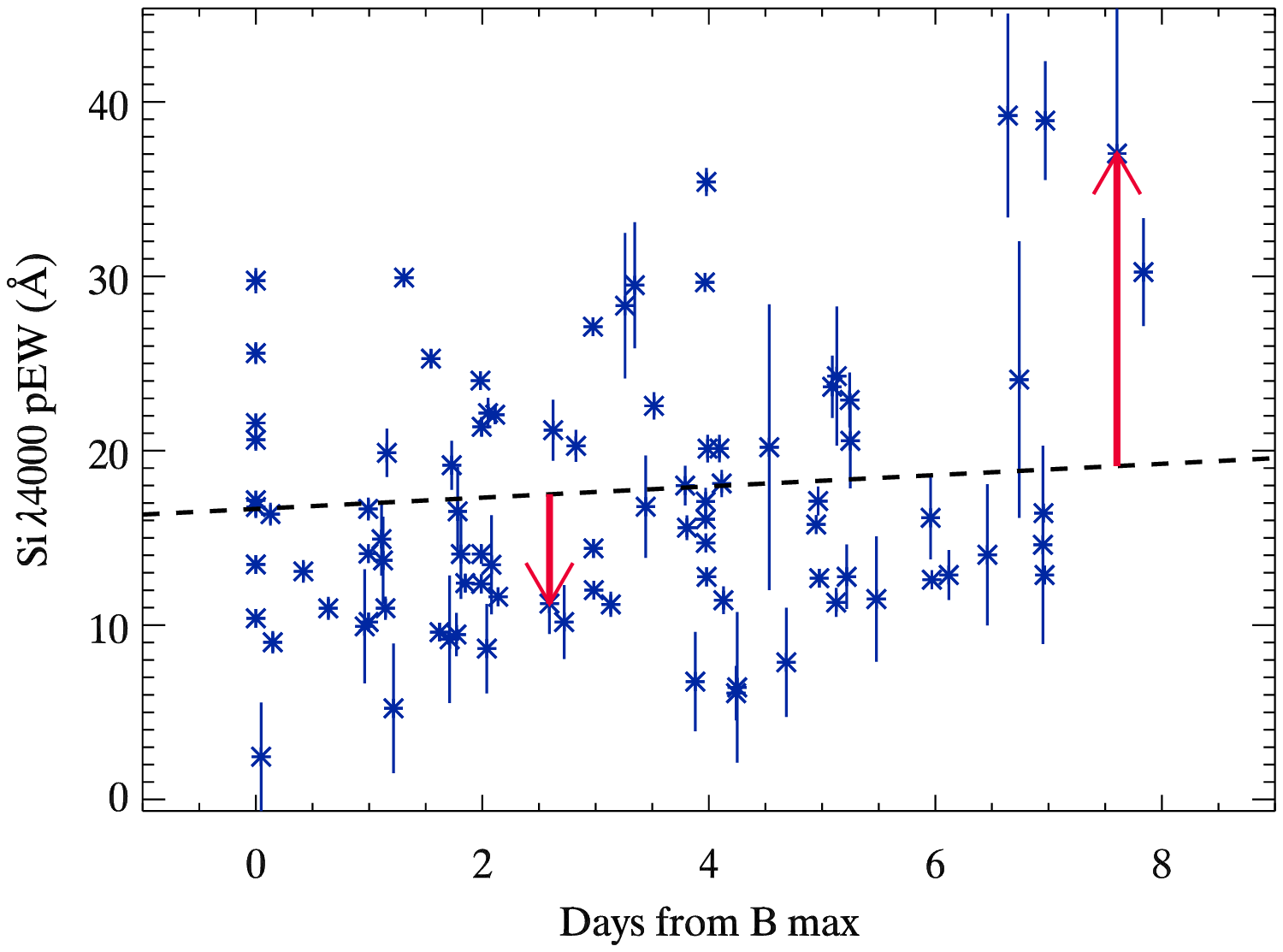}
  \caption{ \refcom{ pEW and $\Delta$pEW measurements of {\si}. Figure shows measurements of {\si} pEW vs. restframe epoch of spectra. The dashed line is the best fit pEW evolution derived using the full sample of N10. $\Delta$pEW is defined as the difference from average evolution for each measurement. Measurements above line will have positive $\Delta$pEW, otherwise negative. Red arrows exemplify one positive and one negative $\Delta$pEW value. }
%The lower panel shows $\Delta$pEW measurements after subtraction of the mean epoch evolution.}
    %\refcom{SDSS NTT/NOT measurements are shown as (black) crosses, CfA as (magenta) stars, SUSPECT as (cyan) boxes and ...} 
}
  \label{fig:deltasample}
\end{figure}

For SNe with multiple spectra in the epoch range, the mean $\Delta$pEW and epoch are used, see Table 2. These steps produce {\si} pEW
measurements of \samplenbr~SNe, \sampleref~from the local sample,
\samplesdss~from Sloan Digital Sky Survey (SDSS), and \samplesnls~from SNLS. \new{The original SDSS and SNLS SN samples are mainly reduced, and in roughly equal fractions, through light curve demands, signal-to-noise requirements, or having no spectra in the studied epoch range.}

Of our sample of {\samplenbr} SNe, \samplegrad~objects have three or more spectra
in the rest-frame epoch range $-7$ to $+20$ (the range where the \sisix~velocity typically is well defined). For these SNe we calculate
the {\sisix} velocity gradient, a linear fit of velocity decrease
per \changed{unit time (km s$^{-1}$ day$^{-1}$)}. We estimate the
uncertainty of these measurements through randomly shifting the
velocity measurements (according to errors) and then recalculate the
gradient. We use the dispersion among these scrambled gradients to
estimate the velocity gradient uncertainty.

See Table 2 for a summary of the data sample.

\section{{\si} and \salttwo~lightcurve color}
\label{sec:si}

\begin{figure}[htbp]
  \centering
  \includegraphics[width=0.7\columnwidth,angle=-90]{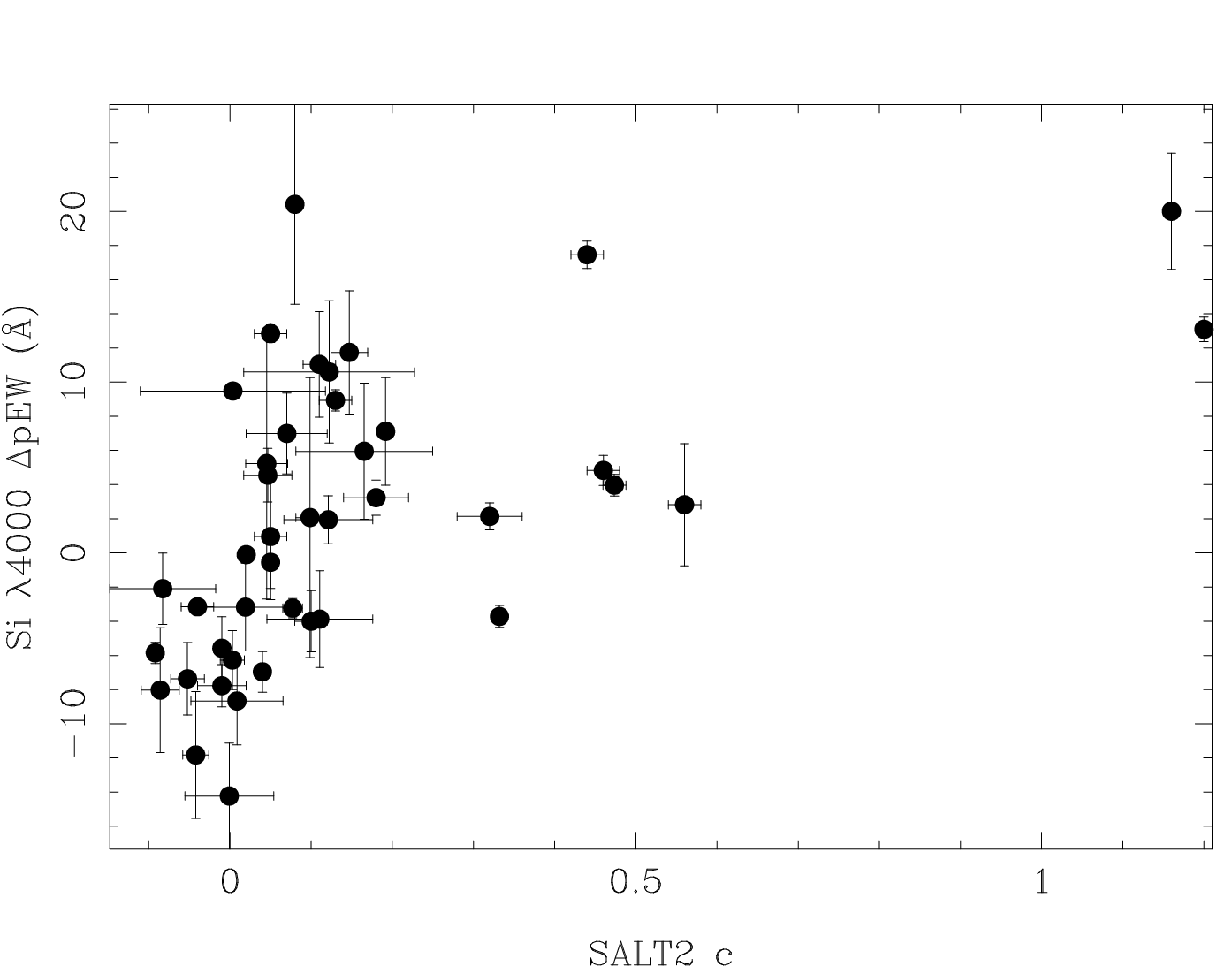}
  \caption{ {\it {\si} pEW \new{difference} ({\dpew}) vs. \salttwo~c.} 
  }
  \label{fig:pewlogc}
\end{figure}

Figure~\ref{fig:pewlogc} shows the comparison of {\si} $\Delta$pEW
measurements for spectra from peak brightness up to \new{+8} days with
\salttwo~color fits. We find that SNe with smaller pEW widths have bluer
light curve colors. This result could be a continuous correlation or
signify different subclasses. The Spearman correlation coefficient for
the correlation is {\sicorrall}, which for the {\samplenbr} objects
corresponds to a rejection of the non-correlation hypothesis at the
\ccorr $\sigma$ level. \changed{If the most reddened SNe are removed
($c>0.3$), the significance changes to \noredccorr~$\sigma$. Recently, several
groups have discussed whether highly reddened objects are
affected by an additional, different, source of extinction
\citep{2010AJ....139..120F,2011A&A...526A.119N,2010arXiv1011.4517F}.}
The significance of the found correlation was also studied through a
simple Monte Carlo test: the pEW values were shuffled so that all color measurements were assigned a random pEW measurement. None of 1000 such iterations yielded a correlation as strong as we find in the data.

\begin{figure}[htbp]
  \centering
  \includegraphics[width=0.6\columnwidth,angle=-90]{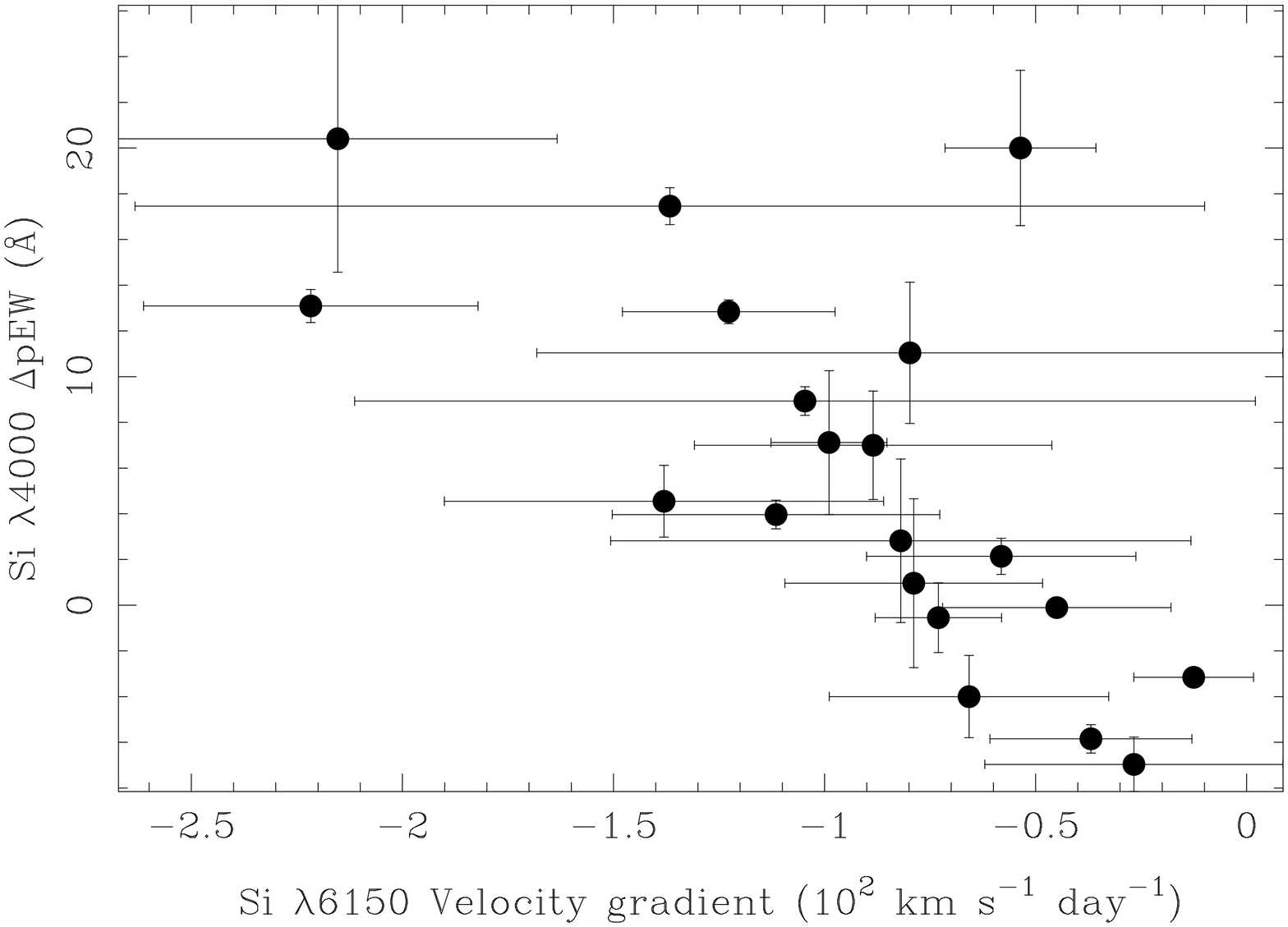}
  \caption{ {\it {\si} {\dpew} vs. velocity gradient of {\sisix}}.}
  \label{fig:pewgrad}
\end{figure}

To further understand the {\si} region of SNe Ia we have studied how {\dpew} measurements vary with other spectral properties.
In Figure~\ref{fig:pewgrad} we compare {\si} {\dpew} measurements
of spectra taken at peak brightness up to 8 days past peak, in rest-frame with {\sisix} velocity gradient. There is a strong correlation where SNe with wider {\si}
features have steeper velocity gradients, thus faster photospheric
velocity evolution.
The Spearman correlation coefficient for the velocity gradient
correlation is $-0.73$, which for the 20 objects corresponds to a
rejection of the non-correlation hypothesis at the \gradcorr $\sigma$
level. Removing \newnew{SNe with \salttwo~$c>0.3$}, the significance becomes
\noredgradcorr~$\sigma$.

\subsection{Systematic Effects}

\begin{figure}[htbp]
  \centering
  \includegraphics[width=0.7\columnwidth,angle=-90]{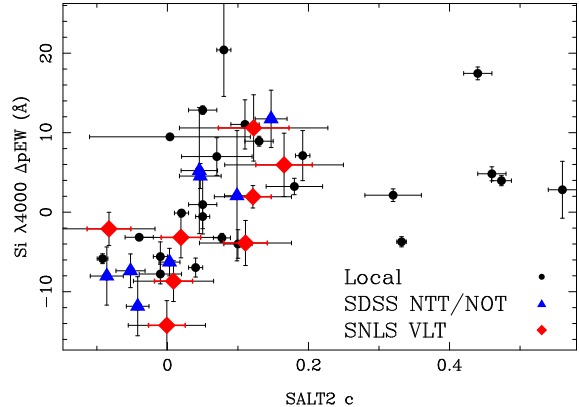}  
  \caption{ {\it {\si} {\dpew} vs. {\salttwo} c with the spectral sample origin displayed.} The bold, red error bars show the original SNLS light curve errors before the uncertainty from the \salttwo~rescaling is applied. \refcom{Note that SN 1999cl and SN 2003cg, both very highly reddened ($c>1$), were omitted from this figure for clarity.}
  }
  \label{fig:peworigin}
\end{figure}

Here we \new{investigate if} the correlation between {\si} width and
\salttwo~color could be caused by systematic effects connected to our
\changed{choice of} data sets or analysis method. In
Figure~\ref{fig:peworigin} we show the information from
Figure~\ref{fig:pewlogc} while highlighting the three different
samples (local, SDSS, and SNLS). These all have different selection
criteria, \new{signal to noise} levels, and host subtraction
methods. Local SNe have high signal to noise and the host galaxy
contamination is, in general, minimal. SDSS and SNLS SNe have lower
signal to noise but are obtained through blind searches. SDSS and SNLS
are host subtracted using different techniques (as described
\linda{in Section~\ref{sec:sample}}). SNLS measurements show errors before and after the uncertainty
from the rescaled {\salttwo} parameters.
None of these subsamples are driving the trends seen. Each sample, if
examined individually, exhibits similar behavior, but at lower
significances. If systematic effects such as noise, sample selection, or
biased host contamination were responsible for the correlation,
samples would not be expected to show the same behavior. See Section 6.4 and Appendices A and B in {\ntio} for a full description of our evaluation of these uncertainties for the SDSS SNe.

\begin{figure}[htbp]
  \centering
  \includegraphics[width=0.7\columnwidth,angle=-90]{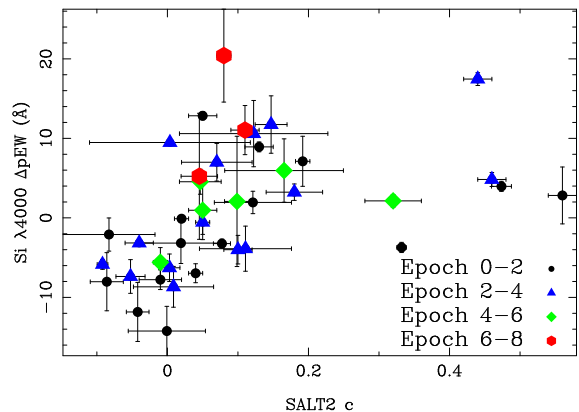}
  \caption{ {\it {\si} {\dpew} vs. {\salttwo} c with epoch information displayed.} \refcom{Note that SN 1999cl and SN 2003cg, both very highly reddened ($c>1$), were omitted from this figure for clarity.}
  }
  \label{fig:pewepoch}
\end{figure}

In a similar way we examine effects from the epoch selection in
Figure~\ref{fig:pewepoch}. We find that the correlation cannot be
explained by epoch differences. 
In Table 1 we show how the Spearman Rank Correlation coefficient changes with different epoch and sample cuts. The significance of the correlation, expressed in standard deviations from random scatter, is given in parenthesis. For the \emph{Low cont} set all SDSS/SNLS SN spectra with estimated host contamination equal to or larger than $20\%$ have been removed. We find that if the epoch range is reduced or the sample restricted the detection $\sigma$ decreases consistently with the reduced number of SNe. \comreply{We note that both N10 and \citet{2011A&A...526A..81B} report no strong correlation between {\si} width and light curve color for epochs close to peak ($-3$ to $3$ days around peak).}

\begin{table}
\caption{Spearman Rank Correlation Coefficients}
% title of Table
\centering
% used for centering table
\begin{tabular}{l c c c c c}
\hline\hline
%inserts double horizontal lines
Ep. & All & Local & SDSS & SNLS & Low Cont \\
% inserts table
%heading
\hline
0-8 & 0.67 (5.8) & 0.54 (2.9) & 0.79 (3.7) & 0.51 (1.8) & 0.65 (4.9) \\ 
0-6 & 0.70 (5.8) & 0.54 (2.7) & 0.81 (3.6) & 0.62 (2.1) & 0.67 (4.8) \\ 
3-8 & 0.71 (4.2) & 0.77 (2.7) & 0.62 (1.6) & 0.50 (1.1) & 0.60 (2.5) \\ 
3-6 & 0.84 (4.7) & 0.96 (3.8) & 0.77 (1.8) & 0.60 (1.0) & 0.72 (2.7) \\ 
\hline
%inserts single line
\end{tabular}
\label{tab:corr}
% is used to refer this table in the text
\end{table}

In {\ntio} we showed that dust reddening of SN templates has negligible impact on pEWs (See Figure 2 in \ntio). That study was extended to include combined effects of reddening and noise: we have constructed Monte Carlo tests where randomized reddening and noise are applied to template SN Ia spectra \citep{2007ApJ...663.1187H}. For extinction, we assume a reddening law
as in \citet{cardelli89} with $R_V = 2.1$. The noise is assumed to be
Gaussian up to the levels where no measurements can be performed. As
with the real sample, we use templates between maximum brightness and
8 days past peak and then subtract the epoch dependence.
We find no trend similar to the one observed, for random distributions of $E(B-V)$, noise and spectral epochs.

Since we only investigated the {\si} pEW correlation with SALT (in
\ntio) and SALT2 color, we cannot exclude that the effect is related
to residual effects from the light curve fitters. Further tests should
also include comparisons with other light curve fitters like
MLCS2k2~\citep{2007ApJ...659..122J} or
SiFTo~\citep{2008ApJ...681..482C}.\footnote{We have repeated the study
with the very small number of SNe with MLCS2k2 or SiFTo
fits. We were unable to make any statistically significant conclusions.
See \ntio~for a more detailed discussion of SALT(1) and MLCS2k2.}
 
Finally, we have investigated whether the measurement pipeline could
create a bias: manual measurements were made on the lower
signal-to-noise spectra that were rejected during the automatic
measurement process. These show an increased scatter, but the trend
does not disappear.

In summary, we cannot find any systematic effect that would explain the
correlation between {\si} pEW  and SALT2 color in Figure~\ref{fig:pewlogc}.

\section{Discussion}
\label{sec:dis}

%% Introduce importance of intrinsic correlation
A definitive detection of a correlation between a spectral feature such as {\si} and light curve color would imply that at least part of the observed \new{color differences} of SNe Ia \new{are intrinsic to the SNe themselves.}
%originates either in the explosion or in the immediate \new{SN} surroundings. 
This intrinsic reddening is connected to the shape and
change of the silicon features of SNe Ia. 
\newnew{At the same time we expect additional reddening effects, at some level, from (at least) host galaxy dust.}
%{\si} width, Branch-types and velocity gradients all probe this.
With two or more different reddening mechanisms, one of which is
intrinsic, it is likely that SN light curve fitters could yield
better cosmological constraints if these mechanisms could be taken into
account separately.

\begin{figure}[htbp]
  \centering
  \includegraphics[width=0.7\columnwidth,angle=-90]{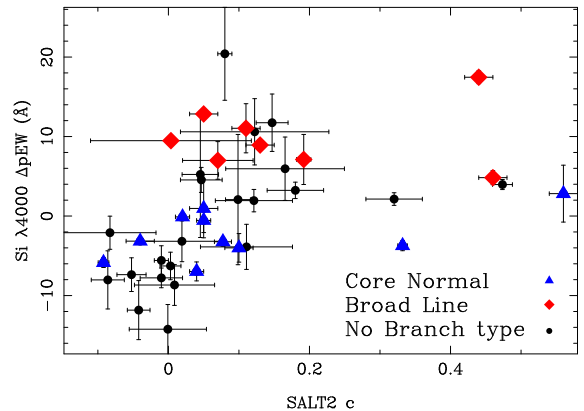}
  \caption{ {\it {\si} {\dpew} vs. {\salttwo} c with the Branch classes marked using different symbols.} \refcom{Note that SN 1999cl and SN 2003cg, both very highly reddened ($c>1$), were omitted from this figure for clarity.}
  }
  \label{fig:pewbranch}
\end{figure}

%% The Branch classification as an old method for subdefinition
Besides a possible direct connection to reddening, the results
presented here can be studied in the light of recent attempts at
finding subtypes of SNe Ia, e.g., the qualitative subtyping scheme 
introduced by \citet{2006PASP..118..560B}. In Figure~\ref{fig:pewbranch} we show {\si} {\dpew} against SALT2 color, but with the Branch classification
\citep[from][]{2006PASP..118..560B,2008PASP..120..135B} marked. Among
the \linda{20} SNe with determined Branch type, all were classified as
either CN or BL. There are no Cool or
SS SNe, consistent with a sample only consisting of
normal SNe Ia.\footnote{The higher redshift samples, SDSS and SNLS,
could contain some ``hidden'' Shallow Silicon SNe that have not been
detected.} Among the \linda{14} moderately reddened ($c \lesssim 0.3$)
SNe with Branch-typing, all SNe with wide {\si} ($\Delta$pEW$\gtrsim \sampledivide$ {\AA}) are BL SNe, while all with
narrow {\si} ($\Delta$pEW$\lesssim \sampledivide$ {\AA}) are CN.
\citet{2008AA...492..535A} found a similar relation between \si~pEW and Branch classification using a wavelet transform technique.

We can further connect {\si} measurements (or thereby Branch
types) to several recent claims regarding further standardization of
SNe: \citet{2009arXiv0905.0340B} and \citet{2011A&A...526A..81B} use
flux ratios, with one ``leg'' in the $4000$-$4500$ {\AA} region, to
standardize low redshift SNe. 
 \comreply{\citet{2009ApJ...699L.139W} use a sample divided according to high or low {\sisix} velocity to obtain either different extinction laws or different intrinsic colors. \citet{2010arXiv1011.4517F}, using the same sample, argue for the later explanation and a connection with the \citet{2007ApJ...662..459K} asymmetric explosion model.}
In Figure~\ref{fig:pewgrad}
we showed how {\si} {\dpew} correlates with {\sisix} velocity
gradient. Since {\sisix} velocity and velocity gradient are related
\citep{2009ApJ...699L.139W}, a connection between {\si} width and
{\sisix} velocity selected samples is also likely.\footnote{We also
see a relation between the {\sisix} velocity around maximum brightness
and the {\si} pEW values, but the correlation with velocity gradient
is stronger.} \comreply{Our results are thus fully consistent with \citet{2009ApJ...699L.139W}.} Exactly how these spectroscopic properties and
subsamples are related, as well as how they are optimally used for SN
Ia cosmology, remains to be determined. 

\begin{figure}[htbp]
  \centering
  \includegraphics[angle=-90,width=0.85\columnwidth]{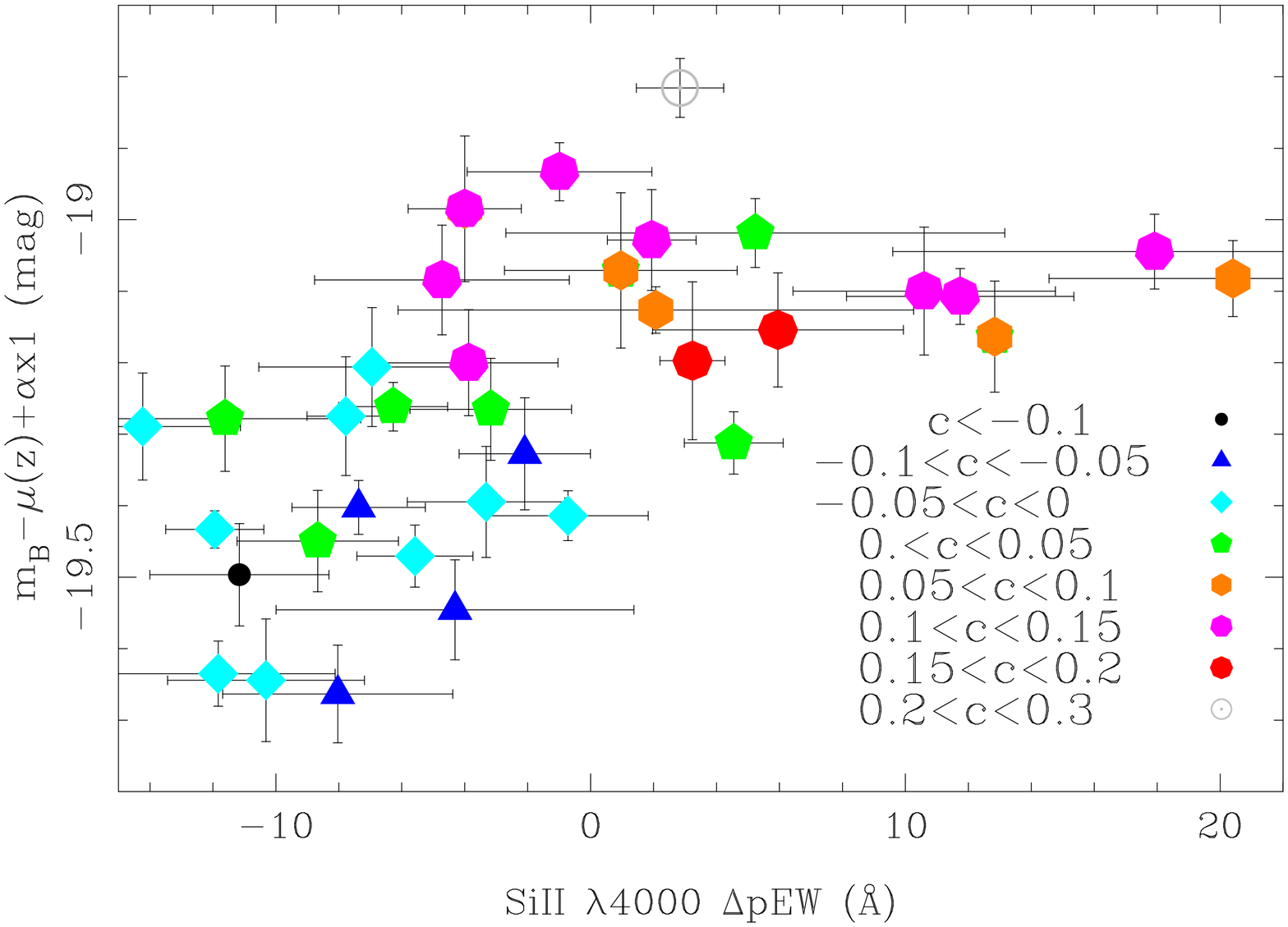}
  \caption{{\it $\Delta$pEW vs distance and light curve width corrected magnitude.} SNe are marked according to \salttwo~c (color). Only SNe in the Hubble flow ($z>0.02$) are included. 
  }
  \label{fig:pewmb}
\end{figure}

%% Guided by these results, see what we can learn about magnitudes etc...
\newnew{We now ask how {\si} {\dpew} measurements relate to SN Ia luminosity.} 
In Figure~\ref{fig:pewmb} we compare {\si} widths
with \linda{absolute} peak magnitude (rest-frame $B$ band). SNe not in the Hubble flow
($z<0.02$) are not included.\footnote{\newnew{Note that all very reddened SNe are removed by this cut.}} We have subtracted the distance modulus, $\mu (z)$, from all magnitudes, assuming a flat $\Lambda$CDM cosmology with
$\Omega_M=0.27$. We have also corrected the magnitudes for light curve width, $x_1$, assuming the $\alpha$ best fit value from Union2 ($\alpha = 0.121$).

The data are sparse, but it is possible that two different behaviors exist:
\newnew{SNe with narrow {\si} regions exhibit a pEW -- magnitude correlation, which could be alternatively described as a reddening -- magnitude correlation.}
SNe with large {\dpew} ($\Delta$pEW$\gsim~\sampledivide$~{\AA}) show a small magnitude scatter that is not improved through light curve color
corrections.

\newnew{Guided by the Branch type we now study the effects of dividing
the sample at $\sim 3$ {\AA}. For the SNe with $\Delta$pEW$\gtrsim
\sampledivide$ {\AA}, the magnitude scatter (without any light curve corrections) is 0.15. Fitting a light curve width correction
to these SNe reduces the scatter to 0.06.} \linda{Photometric
light curve fitters typically produce magnitude scatters around 0.15
mag after light curve corrections.} However, \linda{our}
\newnew{subset consists of only nine objects, including SNLS SNe up 
to $z \sim 0.6$. While this result is
intriguing, more data are needed for a robust
confirmation. }

%% Host galaxy correlations
\newnew{The subsets discussed here could be related to differences between SNe
Ia in different host galaxy environments. Recently \citet{2010arXiv1003.5119S} and \citet{2010arXiv1005.4687L} reported that SNe in passive galaxies are intrinsically brighter after light curve corrections. There are also tentative signs that SNe in passive hosts prefer a less steep extinction law. }
\newnew{Unfortunately we only have secure host galaxy information for a minority of these SNe, and existing information was obtained using different methods. For low redshift SNe with host galaxy information in the literature we note that three out of 16 SNe with narrow {\si} ($\Delta$pEW$< \sampledivide$ {\AA}) occur in E/S0 galaxies, while none out of 8 SNe with wide {\si} ($\Delta$pEW$> \sampledivide$ {\AA}) occur in E/S0 hosts. For the SDSS sample we found in N10 that hosts of SNe with narrow {\si} widths have high specific star formation rate. A coherent analysis of host galaxy properties for our sample is clearly needed. We note that, based on Figure~\ref{fig:pewmb}, SNe with narrow or wide {\si} features would prefer different extinction laws and, possibly, different intrinsic brightness.}
%Larger statistical samples are needed to determine whether this result arises from statistical fluctuations, selection effects or is connected to the intrinsic differences between SNe in different host galaxy types.

%% Nature of Si (done)
\new{The region around 4000 {\AA} is dominated by Si{\sc{ii}}
absorption in normal SNe. However, high signal-to-noise spectra also
exhibit smaller absorption features, attributed to C{\sc{ii}}, Cr{\sc{ii}}, Co{\sc{ii}} or Fe{\sc{iii}} \citep[see,
e.g.][]{2005AJ....130.2278G,2008PASP..120..135B,2010ApJ...713.1073S}. In
spectra with more noise, such small features are not individually
resolved.}  \new{The {\si} feature is thus a complex spectral region,
where absorption from many different ions might be superimposed. The
results presented here need not imply different silicon
abundances. Instead, the presence of other elements
could cause the measured feature width to change. In N10 we examined
composite spectra constructed from SNe with high/low {\si} widths. The
differences we found could be explained through small additional
absorption features. Detailed modeling of this region is needed to
fully understand the {\si} absorption pattern.}

%% Observational remarks
We conclude with two observational remarks. First, the {\si}
feature is sufficiently blue to be observable in optical bands up to
high redshift while the signal-to-noise and resolution requirements for pEW measurements are only moderate (\newnew{S/N $\sim 12$ in $25$ {\AA} bins, roughly a factor of two less than what flux ratios require}). \refcom{Second, measurements of feature widths do not demand spectrophotometric data, only that the host galaxy contamination is moderate (less than $60$\%, see Appendix A in N10).  Provided this can be achieved measurements are thus feasible also with multi-fiber observations. }

The SNLS spectra already included in this study prove that
{\si} width measurements on individual high-redshift spectra are
possible. For the high redshift ($z \sim 0.5$) SNLS spectra used in
this study, observed with VLT FORS1, exposure times of between 40 and
60 minutes were used. Current generation ground-based instruments
(e.g., FORS2 or XShooter) are already sensitive enough to make
accurate measurements of this spectral indicator out to $z \sim 1$. A future
space-based mission (WFIRST) equipped with a slit spectrograph (or IFU) would
be able to provide spectral indicator measurements even beyond $z\sim 1$.

If the correlations between light curve color and feature width
discussed here can be confirmed this would thus constitute a viable
alternative to multi-band photometry (where heavily extincted objects
would be removed using the overall spectral shape).

\section{Conclusions}
\label{sec:conc}

We present evidence of a correlation between \salttwo~color and {\si}
pEW of SN Ia spectra observed between
$B$-band maximum and $8$ days later.
Normal SNe Ia with weak pEW have lower SALT2 $c$ values, while SNe
with wide {\si} absorptions have higher $c$ (are ``redder''). The
Spearman correlation coefficient for the correlation is at the $\ccorr
\sigma$ level. We have looked for spurious effects that could produce the
observed correlation, but have been unable to identify any.

We can further tie the {\si} width to previously defined spectroscopic
subclasses: SNe with wide/deep {\si} are generally of the ``BL''
class with large velocity gradients. SNe with narrow {\si} are ``CN'' with small velocity gradients. \linda{We also show that for our sample of SNe Ia, distance estimates can be improved using {\si} measurements.} More data are needed to study any relationship to host galaxy properties and
extinction. {\si} pEW measurements can, in principle, be done for
high redshift SN spectra with reasonable signal-to-noise.

Our understanding of SNe Ia, and their use as standard
candles, steadily improves. This process will likely allow us to
limit systematic uncertainties and increase the power of SN Ia
cosmology.

\begin{acknowledgements}

We thank Julien Guy for helpful discussions and the anonymous referee for valuable comments.

L.{\"{O}} is partially supported by the Spanish Ministry of Science and Innovation (MICINN) through the Consolider Ingenio-2010 program, under project CSD2007-00060 ``Physics of the Accelerating Universe (PAU)''.

Funding for the Sloan Digital Sky Survey (SDSS) has been provided by
the Alfred P. Sloan Foundation, the Participating Institutions, the
National Aeronautics and Space Administration, the National Science
Foundation, the U.S. Department of Energy, the Japanese
Monbukagakusho, and the Max Planck Society. The SDSS Web site is
http://www.sdss.org/. The SDSS is managed by the Astrophysical
Research Consortium (ARC) for the Participating Institutions. The
Participating Institutions are The University of Chicago, Fermilab,
the Institute for Advanced Study, the Japan Participation Group, The
Johns Hopkins University, Los Alamos National Laboratory, the
Max-Planck-Institute for Astronomy (MPIA), the Max-Planck-Institute
for Astrophysics (MPA), New Mexico State University, University of
Pittsburgh, Princeton University, the United States Naval Observatory,
and the University of Washington.

The paper is partly based on observations made with the Nordic Optical
Telescope, operated on the island of La Palma jointly by Denmark,
Finland, Iceland, Norway, and Sweden, in the Spanish Observatorio del
Roque de los Muchachos of the Instituto de Astrofisica de Canarias.
The data have been taken using ALFOSC, which is owned by the Instituto
de Astrofisica de Andalucia (IAA) and operated at the Nordic Optical
Telescope under agreement between IAA and the NBI.

This paper is partly based on observations collected at  the New Technology Telescope (NTT), operated by  the European Organisation for Astronomical Research in the Southern Hemisphere, Chile. 

The Oskar Klein Centre is funded by the Swedish Research Council.

\end{acknowledgements}

\bibliographystyle{aa}
\bibliography{ms}       % 'publications' is the name of a BibTeX file

\clearpage

\begin{longtable}{p{1.9cm}p{0.9cm}p{1.4cm}p{1.6cm}p{2cm}p{1.4cm}p{3.3cm}p{4cm}} 
\caption{\label{tab:spec} Spectra.} \\ 
\hline \hline 
ID & $z$ & SALT2 $c$ & Si{\sc{ii}} $\lambda 4000$ $\Delta$pEW ({\AA}) & Si{\sc{ii}} $\lambda 6150$ Vel. Grad. ($10^2$ km s$^{-1}$ day$^{-1}$) & Epoch [No. of Spectra] & LC Source & Spectra Source \\ 
\hline 
\endfirsthead 
\caption{continued.}\\ 
\hline\hline 
ID & $z$ & SALT2 $c$ & Si{\sc{ii}} $\lambda 4000$ $\Delta$pEW ({\AA}) & Si{\sc{ii}} $\lambda 6150$ Vel. Grad. ($10^2$ km s$^{-1}$ day$^{-1}$) & Epoch [No. of Spectra] & LC Source & Spectra Source \\ 
\hline 
\endhead 
\hline 
\endfoot 
\endlastfoot 
SN 1989B & 0.0024 & 0.47(0.01) & 3.97(0.63) & $-$1.11(0.39) & 0[1] & \citet{2008AA...492..535A} & \citet{1990AA...237...79B} \\
SN 1990N & 0.0045 & 0.08(0.01) & $-$3.23(0.56) & ... & 1.99[1] & \citet{2008AA...492..535A} & \citet{1993AA...269..423M,1998AJ....115.1096G} \\
SN 1991M & 0.0076 & 0.00(0.11) & 9.48(0.11) & ... & 2.98[1] & \citet{2008AA...492..535A} & \citet{1998AJ....115.1096G} \\
SN 1996X & 0.0078 & 0.02(0.01) & $-$0.10(0.31) & $-$0.45(0.27) & 0.50[2] & \citet{2010ApJ...716..712A} & \citet{2001MNRAS.321..254S} \\
SN 1997do & 0.0105 & 0.11(0.02) & 11.04(3.09) & $-$0.8(0.88) & 7.84[1] & \citet{2010ApJ...716..712A} & \citet{2008AJ....135.1598M} \\
SN 1997dt & 0.0078 & 0.56(0.02) & 2.82(3.58) & $-$0.82(0.69) & 1.12[2] & \citet{2010ApJ...716..712A} & \citet{2008AJ....135.1598M} \\
SN 1998V & 0.0172 & 0.04(0.01) & $-$6.96(1.19) & $-$0.27(0.35) & 1.13[2] & \citet{2010ApJ...716..712A} & \citet{2008AJ....135.1598M} \\
SN 1998aq & 0.0050 & $-$0.09(0.01) & $-$5.85(0.62) & $-$0.37(0.24) & 3.32[15] & \citet{2008AA...492..535A} & \citet{2008AJ....135.1598M,2003AJ....126.1489B} \\
SN 1998bu & 0.0027 & 0.33(0.01) & $-$3.72(0.64) & ... & 0.42[1] & \citet{2008AA...492..535A} & \citet{2008AJ....135.1598M,1999IAUC.7149....2J,2001ApJ...549L.215C,2004AA...426..547S} \\
SN 1998dh & 0.0092 & 0.13(0.02) & 8.93(0.62) & $-$1.05(1.07) & 0[1] & \citet{2010ApJ...716..712A} & \citet{2008AJ....135.1598M} \\
SN 1998dm & 0.0065 & 0.32(0.04) & 2.14(0.79) & $-$0.58(0.32) & 4.09[1] & \citet{2010ApJ...716..712A} & \citet{2008AJ....135.1598M} \\
SN 1998eg & 0.0235 & 0.05(0.02) & 0.96(3.69) & $-$0.79(0.31) & 4.60[2] & \citet{2010ApJ...716..712A} & \citet{2008AJ....135.1598M} \\
SN 1999ao & 0.0544 & 0.08(0.01) & 20.41(5.84) & $-$2.15(0.52) & 6.64[1] & \citet{2010ApJ...716..712A} & \citet{2007AA...470..411G} \\
SN 1999ar & 0.1561 & $-$0.01(0.01) & $-$5.58(1.84) & ... & 5.22[1] & \citet{2010ApJ...716..712A} & \citet{2007AA...470..411G} \\
SN 1999bm & 0.1428 & 0.18(0.04) & 3.24(1.03) & ... & 3.94[2] & \citet{2010ApJ...716..712A} & \citet{2007AA...470..411G} \\
SN 1999bn & 0.1285 & $-$0.01(0.03) & $-$7.78(1.24) & ... & 1.77[1] & \citet{2010ApJ...716..712A} & \citet{2007AA...470..411G} \\
SN 1999cc & 0.0315 & 0.05(0.02) & 12.84(0.52) & $-$1.23(0.25) & 1.31[1] & \citet{2010ApJ...716..712A} & \citet{2008AJ....135.1598M} \\
SN 1999cl & 0.0087 & 1.20(0.01) & 13.09(0.72) & $-$2.22(0.4) & 0[1] & \citet{2010ApJ...716..712A} & \citet{2008AJ....135.1598M} \\
SN 1999ej & 0.0154 & 0.07(0.05) & 7.00(2.37) & $-$0.88(0.42) & 2.53[2] & \citet{2010ApJ...716..712A} & \citet{2008AJ....135.1598M} \\
SN 1999gd & 0.0193 & 0.46(0.02) & 4.83(0.88) & ... & 2.05[1] & \citet{2010ApJ...716..712A} & \citet{2008AJ....135.1598M} \\
SN 2000cf & 0.0365 & 0.04(0.02) & 1.67(1.83) & $-$0.53(0.44) & 3.31[2] & \citet{2010ApJ...716..712A} & \citet{2008AJ....135.1598M} \\
SN 2000fa & 0.0218 & 0.10(0.02) & $-$4.00(1.80) & $-$0.66(0.33) & 2.83[2] & \citet{2010ApJ...716..712A} & \citet{2008AJ....135.1598M} \\
SN 2002bo & 0.0053 & 0.44(0.02) & 17.46(0.80) & $-$1.37(1.27) & 3.98[1] & \citet{2010ApJ...716..712A} & \citet{2004MNRAS.348..261B} \\
SN 2002er & 0.0086 & 0.19(0.01) & 7.12(3.15) & $-$0.99(0.14) & 1.98[3] & \citet{2008AA...492..535A} & \citet{2005AA...436.1021K} \\
SN 2003cg & 0.0041 & 1.16(0.01) & 20.01(3.40) & $-$0.54(0.18) & 6.97[1] & \citet{2010ApJ...716..712A} & \citet{2006MNRAS.369.1880E} \\
SN 2003du & 0.0064 & $-$0.04(0.02) & $-$3.15(0.24) & $-$0.13(0.14) & 2.98[5] & \citet{2010ApJ...716..712A} & \citet{2007AA...469..645S,2005AA...429..667A,2005ASPC..342..250G} \\
SN 2005cf & 0.0070 & 0.05(0.01) & $-$0.55(1.53) & $-$0.73(0.15) & 3.77[5] & \citet{2010ApJ...716..712A} & \citet{2007AA...471..527G,2007AIPC..937..311L} \\
SN 2006fx & 0.2239 & 0.11(0.02) & $-$0.99(2.93) & ... & 3.44[1] & SDSS-II & \citet{2011AA...526A..28O} \\
SN 2006kq & 0.1985 & $-$0.02(0.02) & $-$0.72(2.55) & ... & 1.78[1] & SDSS-II & \citet{2011AA...526A..28O} \\
SN 2006nw & 0.1570 & 0.00(0.01) & $-$6.27(1.73) & ... & 2.59[1] & SDSS-II & \citet{2011AA...526A..28O} \\
SN 2007jd & 0.0727 & 0.23(0.03) & 2.84(1.39) & $-$5.6(2.5) & 1.16[1] & SDSS-II & \citet{2011AA...526A..28O} \\
SN 2007jk & 0.1828 & 0.15(0.02) & 11.74(3.61) & ... & 3.35[1] & SDSS-II & \citet{2011AA...526A..28O} \\
SN 2007ka & 0.2180 & $-$0.05(0.02) & $-$7.37(2.12) & ... & 2.72[1] & SDSS-II & \citet{2011AA...526A..28O} \\
SN 2007lh & 0.1980 & 0.05(0.03) & 5.23(7.93) & ... & 6.74[1] & SDSS-II & \citet{2011AA...526A..28O} \\
SN 2007lm & 0.2130 & 0.10(0.02) & 2.07(8.19) & ... & 4.53[1] & SDSS-II & \citet{2011AA...526A..28O} \\
SN 2007ol & 0.0560 & 0.05(0.03) & 4.55(1.57) & $-$1.38(0.52) & 5.24[1] & SDSS-II & \citet{2011AA...526A..28O} \\
SN 2007pf & $-$1.000 & $-$0.02(0.02) & $-$11.94(1.56) & ... & 4.24[1] & SDSS-II & \citet{2011AA...526A..28O} \\
SN 2007pu & 0.0890 & $-$0.04(0.02) & $-$11.83(3.71) & ... & 1.22[1] & SDSS-II & \citet{2011AA...526A..28O} \\
SN 2007qg & 0.3139 & $-$0.02(0.03) & $-$10.31(3.13) & ... & 4.68[1] & SDSS-II & \citet{2011AA...526A..28O} \\
SN 2007qh & 0.2477 & 0.10(0.02) & 17.91(8.32) & ... & 7.60[1] & SDSS-II & \citet{2011AA...526A..28O} \\
SN 2007ql & $-$1.000 & $-$0.00(0.02) & $-$3.32(2.50) & ... & 1.12[1] & SDSS-II & \citet{2011AA...526A..28O} \\
SN 2007qs & 0.2910 & $-$0.09(0.02) & $-$8.03(3.65) & ... & 1.71[1] & SDSS-II & \citet{2011AA...526A..28O} \\
SNLS 03D4at & $-$1.000 & $-$0.00(0.14) & $-$6.94(3.60) & ... & 5.48[1] & SDSS-II & \citet{2011AA...526A..28O} \\
SNLS 04D1rh & 0.4360 & $-$0.00(0.05) & $-$14.23(3.11) & ... & 0.05[1] & \citet{2010arXiv1010.4743G} & \citet{2009AA...507...85B} \\
SNLS 04D2fp & 0.4150 & 0.02(0.06) & $-$3.17(2.57) & ... & 1.81[1] & \citet{2010arXiv1010.4743G} & \citet{2009AA...507...85B} \\
SNLS 04D2fs & 0.3570 & 0.12(0.05) & 1.94(1.41) & ... & 1.73[1] & \citet{2010arXiv1010.4743G} & \citet{2009AA...507...85B} \\
SNLS 04D2mc & 0.3480 & 0.15(0.07) & $-$4.72(4.05) & ... & 6.46[1] & SDSS-II & \citet{2011AA...526A..28O} \\
SNLS 04D4bq & 0.5500 & 0.17(0.08) & 5.95(3.98) & ... & 5.13[1] & \citet{2010arXiv1010.4743G} & \citet{2009AA...507...85B} \\
SNLS 05D1cb & $-$1.000 & 0.04(0.09) & $-$11.61(4.31) & ... & 4.25[1] & SDSS-II & \citet{2011AA...526A..28O} \\
SNLS 05D2ac & 0.4790 & 0.01(0.06) & $-$8.67(2.56) & ... & 2.04[1] & \citet{2010arXiv1010.4743G} & \citet{2009AA...507...85B} \\
SNLS 05D2dy & 0.5100 & $-$0.08(0.07) & $-$2.09(2.08) & ... & 1.11[1] & \citet{2010arXiv1010.4743G} & \citet{2009AA...507...85B} \\
SNLS 05D4be & $-$1.000 & $-$0.12(0.06) & $-$11.16(2.85) & ... & 3.88[1] & SDSS-II & \citet{2011AA...526A..28O} \\
SNLS 05D4cw & 0.3750 & $-$0.09(0.05) & $-$4.31(5.68) & ... & 6.95[1] & SDSS-II & \citet{2011AA...526A..28O} \\
SNLS 05D4ek & 0.5360 & 0.11(0.07) & $-$3.87(2.84) & ... & 2.08[1] & \citet{2010arXiv1010.4743G} & \citet{2009AA...507...85B} \\
SNLS 06D2cc & 0.5320 & 0.12(0.11) & 10.60(4.17) & ... & 3.26[1] & \citet{2010arXiv1010.4743G} & \citet{2009AA...507...85B} \\
\end{longtable}

\end{document}